# Omnidirectional Transponder for Narrow-band Radar Calibration

Oren Cohen, Moshe Vana
ELTA Systems Ltd, 100 Yitzhak Hanassi Blvd., Ashdod, Israel
Correspondence: orencohen@elta.co.il

## Abstract

Conventional reference targets for Synthetic Aperture Radar (SAR) calibration, such as corner reflectors and standard transponders, are often inherently large and suffer from limited viewing angles. This paper presents a novel frequency-translating transponder architecture that circumvents these limitations, enabling a truly compact, single-antenna design capable of omnidirectional operation. While the operational bandwidth is consequently narrowed, restricting its use primarily to azimuth-direction calibrations, the design excels at providing highly accurate pulse-to-pulse phase measurements across the synthetic aperture. The transponder was prototyped and experimentally validated with a drone-mounted SAR. The results demonstrate the transponder's significant potential for applications requiring omnidirectional reference targets, such as Circular SAR (CSAR) and bistatic SAR.

## 1 Introduction

Reference targets are commonly used to calibrate synthetic aperture radar (SAR) systems. There are two main kinds of reference targets: passive and active. Passive targets [1], such as the commonly used corner reflector, are characterized by their simple structure, low construction cost, and stable performance without the need for an external energy source. In contrast, active reference targets, known as transponders, are more complex in both their design and operation but offer several distinct advantages over their passive counterparts, with the primary benefits being their relatively compact size, the ability to generate a large and tunable Radar Cross Section (RCS), and their support for polarimetric calibration.

A basic transponder design consists of a receiving (Rx) antenna, an amplifier, and a transmitting (Tx) antenna [2], as schematically shown in **Figure 1a**. The RCS of the transponder is directly proportional to the gain of its internal amplifier; it is therefore beneficial to use the highest gain possible to generate a strong and easily detectable signal. However, the isolation between the Tx and Rx antennas ultimately limits the maximum stable amplifier gain, and consequently the maximum RCS, since for proper operation, this isolation must be greater than the amplifier's gain. If this is not the case, significant signal degradation can occur, including phase and amplitude distortion across the operational frequency band, an increased receiver noise floor, and in the worst-case scenario, self-oscillations that render the device useless. To achieve the high isolation required, one may physically separate the Tx and Rx antennas, increase the directivity of the antennas, or use shielding materials between them [3]. Each of these solutions introduces a trade-off. For instance, physical separation considerably increases the overall size of the transponder, counteracting one of its primary advantages. Similarly, employing antennas with increased directivity to reduce mutual coupling inherently makes the transponder more directional, limiting its effective operation to a narrow range of viewing angles, much like a corner reflector.

In this paper, we explore a transponder architecture that circumvents the isolation problem by translating the signal's frequency band before re-transmitting it. This scheme can be successfully realized for radars employing either Matched Filter (MF) or Frequency Modulated Continuous Wave (FMCW) operation. In both architectures, a key consideration is to prevent interference between the transponder's echo and the surrounding clutter return. This is achieved by assigning a dedicated portion of the radar receiver's Intermediate Frequency (IF) spectrum exclusively to the returned transponder signal. For a system using a Matched Filter, this can be implemented by limiting the transmitted radar bandwidth to be narrower than the total available IF bandwidth, thus leaving a vacant frequency band into which the transponder echo can be shifted. In the case of an FMCW system, the frequency shift can be designed to place the transponder's echo at a virtual far range, where the energy from clutter returns is typically low.

Separating the receive and transmit frequencies offers significant advantages over the basic transponder design, primarily by enabling the use of a single antenna for both functions. This single-antenna operation provides two key benefits. First, it removes any constraints on the antenna's directivity that were previously imposed to achieve isolation; an omnidirectional antenna can even be used, allowing the transponder to receive and return radar signals from all directions. Second, this architecture allows for a considerable size reduction relative to a conventional transponder. With the need for physical separation between two antennas eliminated, the transponder's physical footprint is reduced to the size of the single antenna itself and its associated electronics.

This approach, however, comes at the expense of a reduced operational bandwidth, which consequently limits its calibration applications. Since SAR imaging is inherently wideband in nature, this reduced bandwidth primarily restricts the transponder's utility to calibrations related to the azimuth direction. One relevant use case in this sense is range estimation. By deploying at least three such transponders within the imaging scene, their range measurements can be used to fully determine the radar's location in scenarios where the on-board Inertial Navigation System (INS) accuracy is insufficient. This capability can be especially useful for focusing circular SAR (CSAR) images in experimental settings [4]. Since the flight trajectory is circular, omnidirectional reference targets are needed, a requirement this transponder design can meet. Furthermore, as CSAR is inherently more sensitive to INS errors, the precise positional data derived from the transponder can be used to mitigate these errors and improve image quality.

Beyond monostatic calibration, the proposed transponder offers significant utility as a reference target for bistatic radar systems. In bistatic configurations, conventional passive targets such as corner reflectors are generally only effective in quasi-monostatic scenarios due to their inherently narrow beamwidths. Conversely, the omnidirectional capability of the developed transponder allows it to be utilized across arbitrary bistatic geometries.

# 2 Design

Naively, a frequency-shifting transponder can theoretically be designed around a single RF mixer, as illustrated in **Figure 1b**. In this configuration, a circulator is used to allow a single antenna to be shared between the receiving and transmitting paths. Upon reception, a Low-Noise Amplifier (LNA) first amplifies the incoming signal. This signal is then fed to a mixer, which shifts its frequency by mixing it with a Local Oscillator (LO) signal. A Band-Pass Filter (BPF) is subsequently used to select the desired frequency-shifted signal and reject unwanted mixer products. Finally, a transmit amplifier boosts the signal's power before it is routed back through the circulator for re-transmission.

However, such a straightforward approach suffers from two significant shortcomings. First, it is difficult to completely isolate the original received frequency from the transmitting path. The isolation, limited by the mixer's inherent port-to-port leakage and the finite rejection of the band-pass filter, directly constrains the transponder's maximum stable gain and achievable RCS, similar to the limitation imposed by antenna coupling in the basic design discussed previously. Second, the transponder's LO is not synchronized with the radar's master clock, which introduces phase noise and a non-deterministic Doppler shift onto the retransmitted signal, rendering it unsuitable for stable and coherent calibration tasks.

To overcome the shortcomings of the naïve single-mixer design, the RF chain can be upgraded to that in **Figure 1c**. This design addresses both the isolation and synchronization issues. First, a superheterodyne-like approach is adopted, where the received signal is down-converted to a low Intermediate Frequency (IF) before being filtered and subsequently up-converted for transmission. Operating at a lower IF makes it significantly easier to implement a Band-Pass Filter (BPF) with a sharp roll-off and achieve far better frequency selectivity compared to the RF BPF used in the naïve design, thus greatly improving isolation. Second, to solve the synchronization problem, the up-conversion stage is modified to generate two tones instead of one. This two-tone scheme allows the radar to synchronize with the transponder, enabling the estimation and correction of both the transponder's additive phase noise and its Doppler shift. This is conceptually similar to the two-tone method often employed in Vector Network Analyzers (VNAs) for conducting phase response measurements of frequency-converting devices with embedded-LO [5].

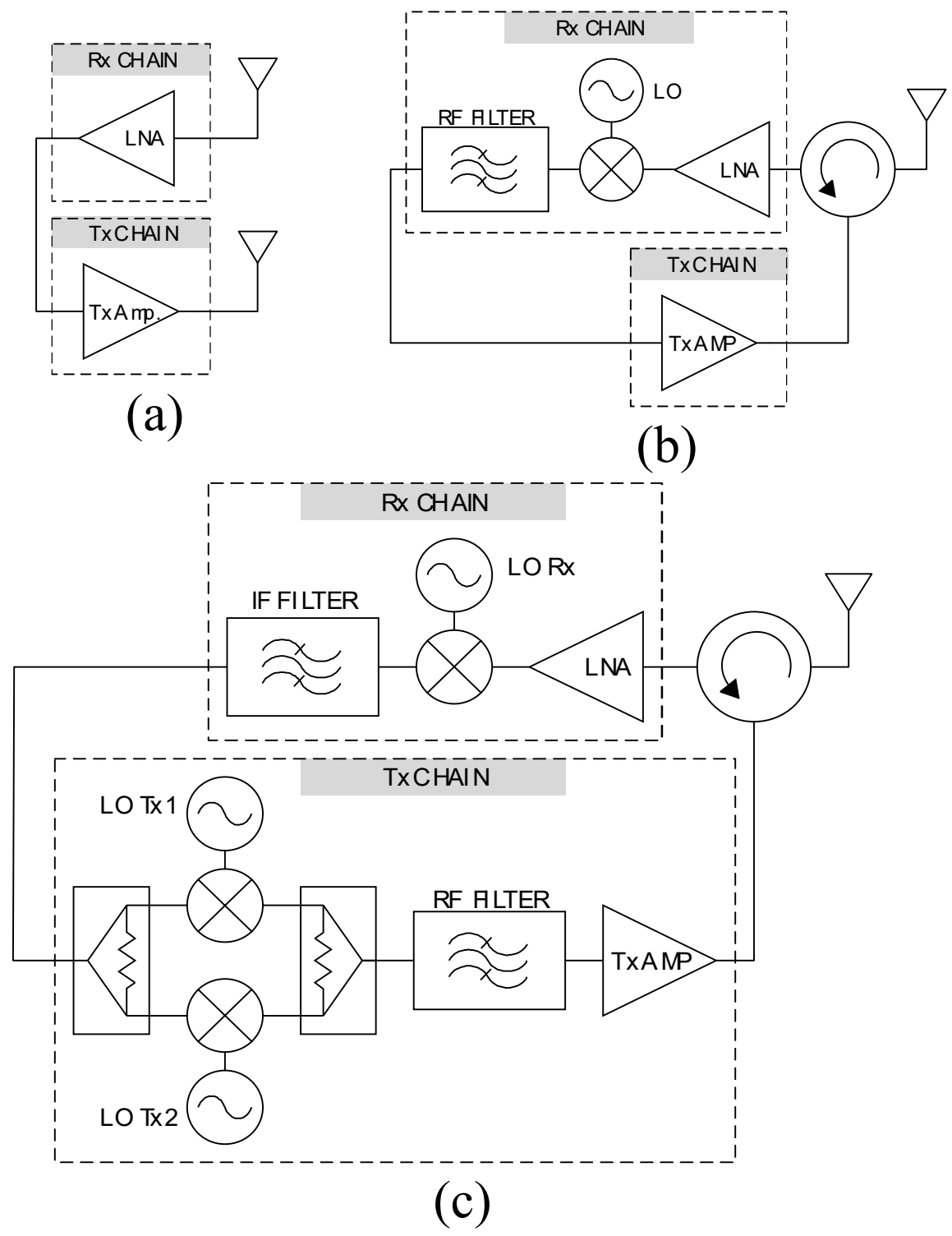

**Figure 1** Block diagrams of three transponder designs, showing the key components in both the Tx and Rx signal paths. (a) A basic transponder. (b) A single conversion frequency shifting transponder. (c) A dual conversion two-tone transponder presented in this paper.

# 3 Method

In this work, a dual-conversion frequency-translating transponder was constructed and tested as a reference target for a drone-mounted X-band FMCW SAR radar. The main focus of this experiment was to evaluate the transponder's range estimation ability and accuracy, and to compare its performance directly to that achieved with a conventional corner reflector.

## 3.1 Hardware

The overall setup and frequency plan are illustrated in **Figure 2**. The radar was operated at a 9.75 GHz RF frequency with a 500 MHz LFM bandwidth ($B$), a 450 us pulse width ($T$), and an IF band that spans from DC to 30 MHz. As the transponder's frequency shift is constrained by this available IF bandwidth, the transponder was configured to shift the input signals by 20 MHz and 25 MHz, ensuring that their corresponding post de-chirp frequencies would fall well within the radar's IF range.

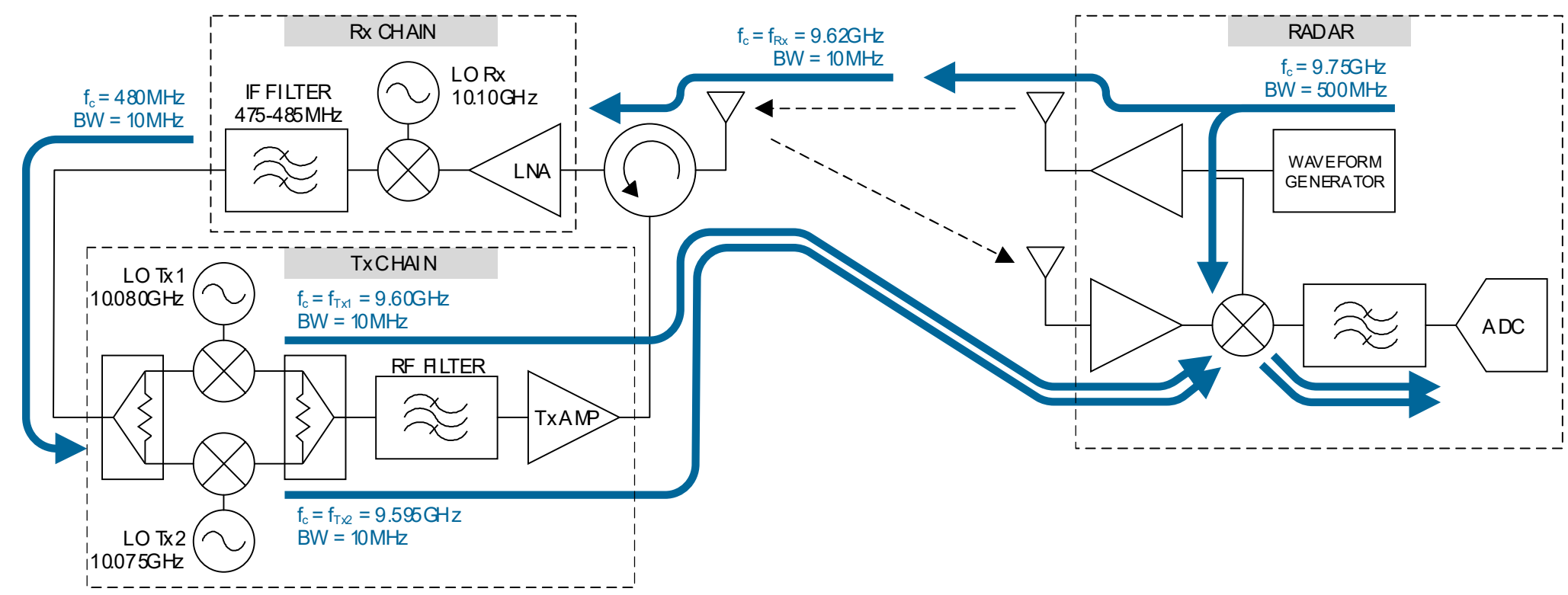

**Figure 2** Block diagrams and frequency plan of both transponder and radar. Notably, out of the full radar bandwidth, the transponder only receives a narrow slice of the spectrum between 9.615 GHz and 9.625 GHz.

To allow the IF filter to effectively reject leakage of the transmitted signal back into the receiving path, the transponder's operational bandwidth must be narrower than its frequency shift. This constraint limited the transponder bandwidth to 10 MHz in our design. The particular IF was then selected to match the center frequency of the sharpest 10 MHz bandpass filter that was readily available. This proved to be a 480 MHz Surface Acoustic Wave (SAW) filter, whose measured frequency response is plotted in **Figure 3**.

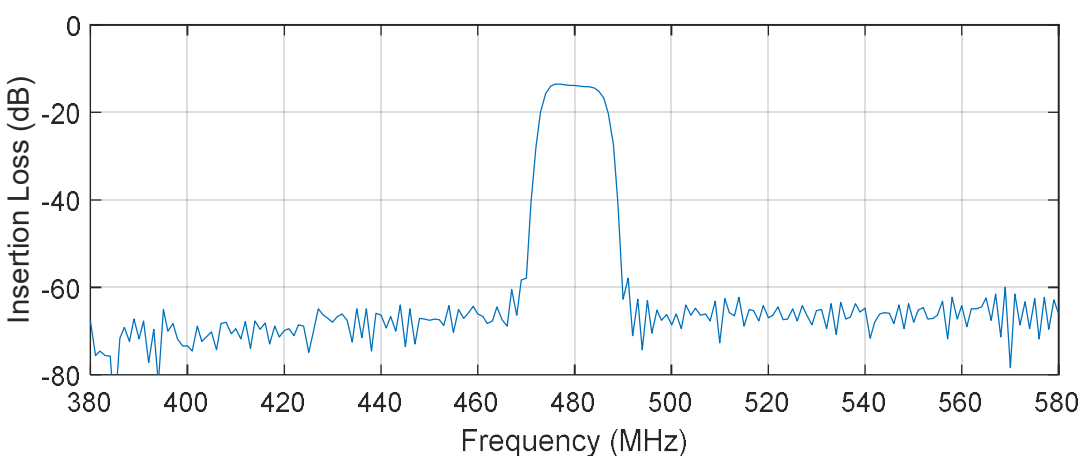

**Figure 3** Insertion loss of the SAW filter used as the IF filter. Two such filters were cascaded and used to achieve better rejection.

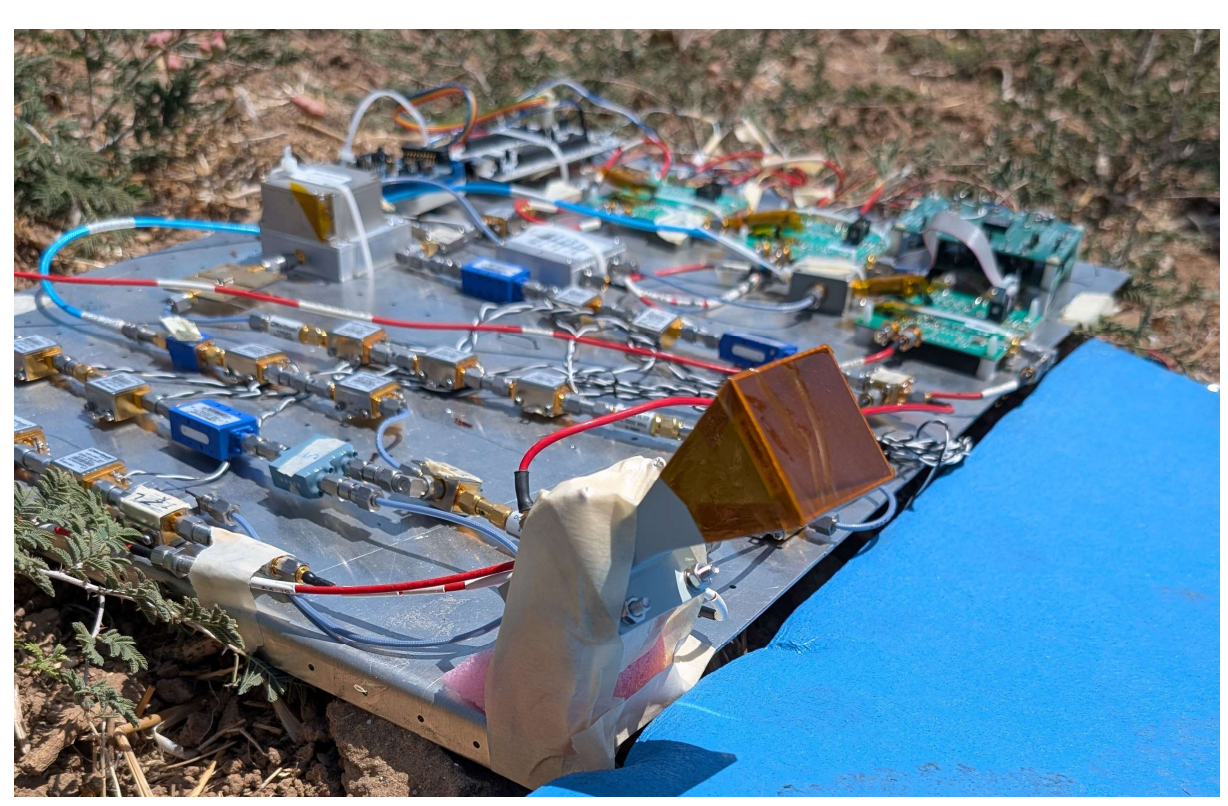

(a)

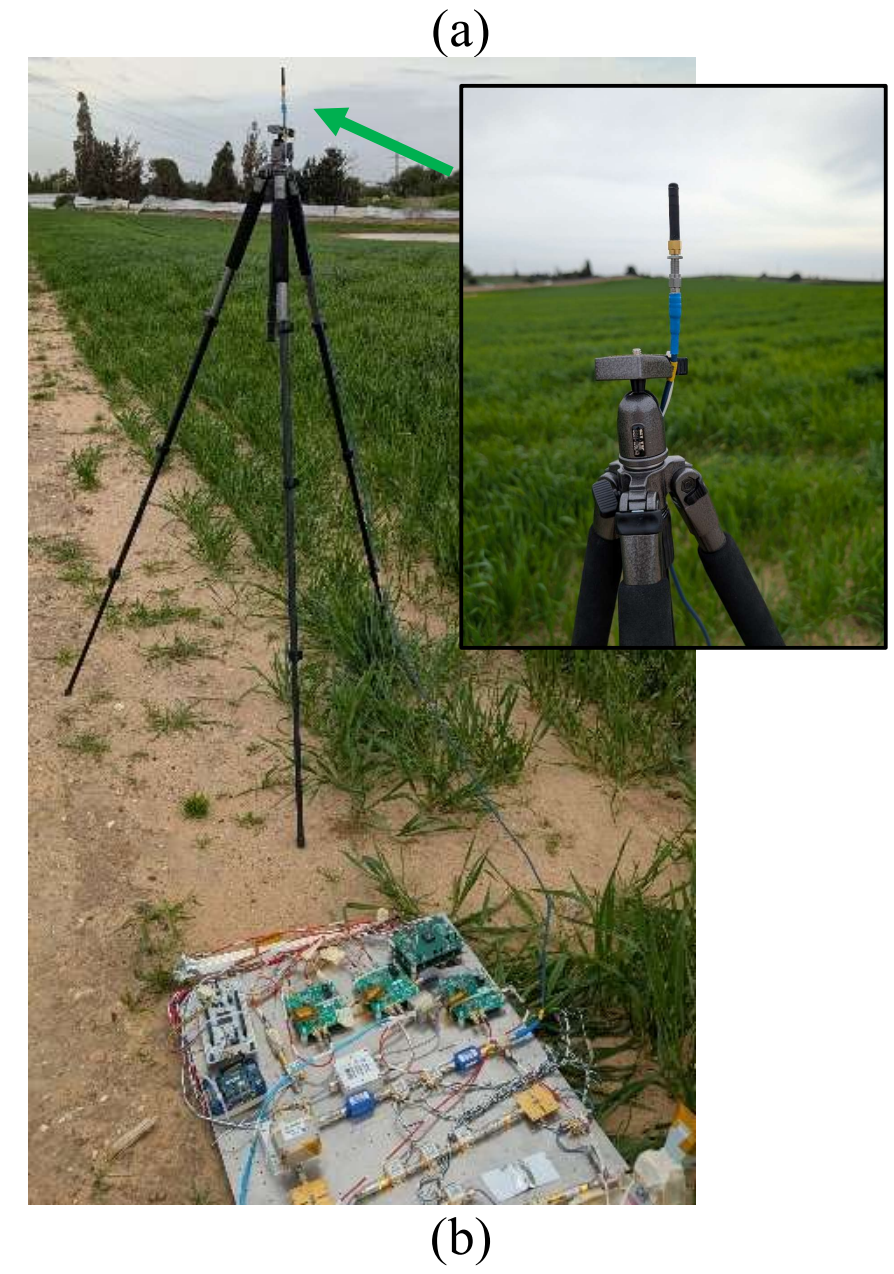

(b)

**Figure 4** Photographs of the transponder equipped with **(a)** a directional horn antenna and **(b)** an omnidirectional dipole antenna.

The three LOs were generated by three separate Phase-Locked Loops (PLLs), all of which were fed by the same clock reference to ensure phase coherence and phase noise correlation.

The system was evaluated using two interchangeable antenna configurations: a standard horn antenna providing 10 dBi of gain and a 55-degree beamwidth, and an omnidirectional dipole antenna with 2 dBi of gain. The transponder's overall RF chain gain was adjusted to 50 dB for the horn configuration and to 55 dB for the dipole.

The transponder was constructed using readily available off-the-shelf (COTS) connectorized RF components, as shown in **Figure 4**. While this modular approach allowed for rapid prototyping and design iterations, it inevitably resulted in a relatively large physical size. Nevertheless, we expect that a production version designed with Surface Mount Technology (SMT) could be easily miniaturized to a compact form factor, approximately the size of a Rubik's Cube.

## 3.2 Analysis

The transponder's range can be estimated from the raw radar Analog-to-Digital Converter (ADC) samples using a relatively straightforward processing. Following the dechirp operation, the transponder echo conveniently appears as two distinct single tones at the ADC input, $exp(i\phi_1(t)) + exp(i\phi_2(t))$. The range information is embedded in the phase history of each of these tones, $\phi_1$ and $\phi_2$, which are mathematically expressed by **Equation 1**, where $R$ is the transponder range, $f_{Rx}$ is the center frequency of the signal received by the transponder, $f_{Tx1}$ and $f_{Tx2}$ are the frequencies of the two tones transmitted by the transponder, $f_{XO}$ is the transponder internal clock reference frequency, and $\phi_n(t)$ is this clock phase noise.

$$\phi_1(t) = f_{Rx}\frac{2\pi R}{c} + f_{Tx1}\frac{2\pi R}{c} - 2\pi(f_{Tx1} - f_{Rx})t + \frac{f_{Tx1} - f_{Rx}}{f_{XO}}\phi_n(t) + \frac{B}{T}\frac{4\pi R}{c}t - \frac{\pi B}{T}\left(\frac{2R}{c}\right)^2$$

$$\phi_2(t) = f_{Rx}\frac{2\pi R}{c} + f_{Tx2}\frac{2\pi R}{c} - 2\pi(f_{Tx2} - f_{Rx})t + \frac{f_{Tx2} - f_{Rx}}{f_{XO}}\phi_n(t) + \frac{B}{T}\frac{4\pi R}{c}t - \frac{\pi B}{T}\left(\frac{2R}{c}\right)^2$$

**Equation 1**

The terms within each expression in **Equation 1** correspond to specific physical effects. The first two terms represent the phase accumulated due to the signal's time of flight between the radar and the transponder. The third term is the frequency shift introduced by the transponder, while the fourth term accounts for the transponder's residual phase noise. The fifth term is the characteristic range-dependent linear phase inherent to FMCW radars. The final term represents the residual video phase (RVP).

In the expressions presented in **Equation 1**, all hardware delays inherent in both the radar and transponder systems—such as those from cables and filters—are assumed to be zero to maintain analytical simplicity. In practice, however, these delays must be accounted for and compensated to achieve proper results.

The transponder's range is estimated in a two-step process. First, an FFT is performed on the received signal. Next, a peak search on the resulting spectrum identifies the two tones, allowing us to extract their frequencies ($f_1$, $f_2$) and initial phases ($\phi_1(0)$, $\phi_2(0)$).

The range between the radar and transponder can be estimated independently using one of the following schemes:

- **Scheme 1 - Absolute Range**: The absolute range ($R_{abs}$) is calculated from the tone frequencies ($f_1$, $f_2$) according to **Equation 2**, where it is assumed that the transponder's phase noise is stationary between consecutive pulses (i.e., $d\phi_n(t)/dt = 0$).
- **Scheme 2 - Relative Range Change**: The relative change in range from pulse to pulse is estimated using the extracted phases as described in **Equation 3**. This calculation requires the quadratic RVP term, which can be estimated from the absolute range measurement. Due to the phase wrapping every 2π, this method directly yields only the incremental range change, necessitating a phase unwrapping to reconstruct the full relative trajectory ($R_{rel}$). However, this phase-based approach provides a significantly more accurate range change measurement compared to the absolute estimation scheme.

$$R_{abs} = \frac{cT}{2B}(f_1 + f_2 - f_{Rx} + f_{Tx1} - f_{Rx} + f_{Tx2})$$

**Equation 2**

$$R_{rel} = \frac{c}{4\pi f_{Rx}}\left(\frac{\phi_1(0) - \frac{f_{Tx1} - f_{Rx}}{f_{Tx2} - f_{Rx}}\phi_2(0)}{1 - \frac{f_{Tx1} - f_{Rx}}{f_{Tx2} - f_{Rx}}} + \frac{\pi B}{T}\left(\frac{2R_{abs}}{c}\right)^2\right)$$

**Equation 3**

# 4 Experimental Results

To validate the performance of the proposed transponder, two field experiments were conducted using a drone-mounted, X-band FMCW SAR system. In the first experiment, the drone flew along a linear trajectory, while in the second, it executed a circular flight path around the target.

## 4.1 Linear Flight Experiment

For this experiment, the transponder was equipped with the directional horn antenna. The drone flew along a straight 260m trajectory at an altitude of approximately 100 meters above ground level, with the transponder positioned on the ground at a range of approximately 280 meters from the flight path's ground center. The flight geometry is illustrated in **Figure 5**. Overall, the transponder was observed across a ±25° range of viewing angles. For a direct performance comparison, two conventional corner reflectors with an edge length of 0.85 meters were also placed within the imaging scene in the vicinity of the transponder.

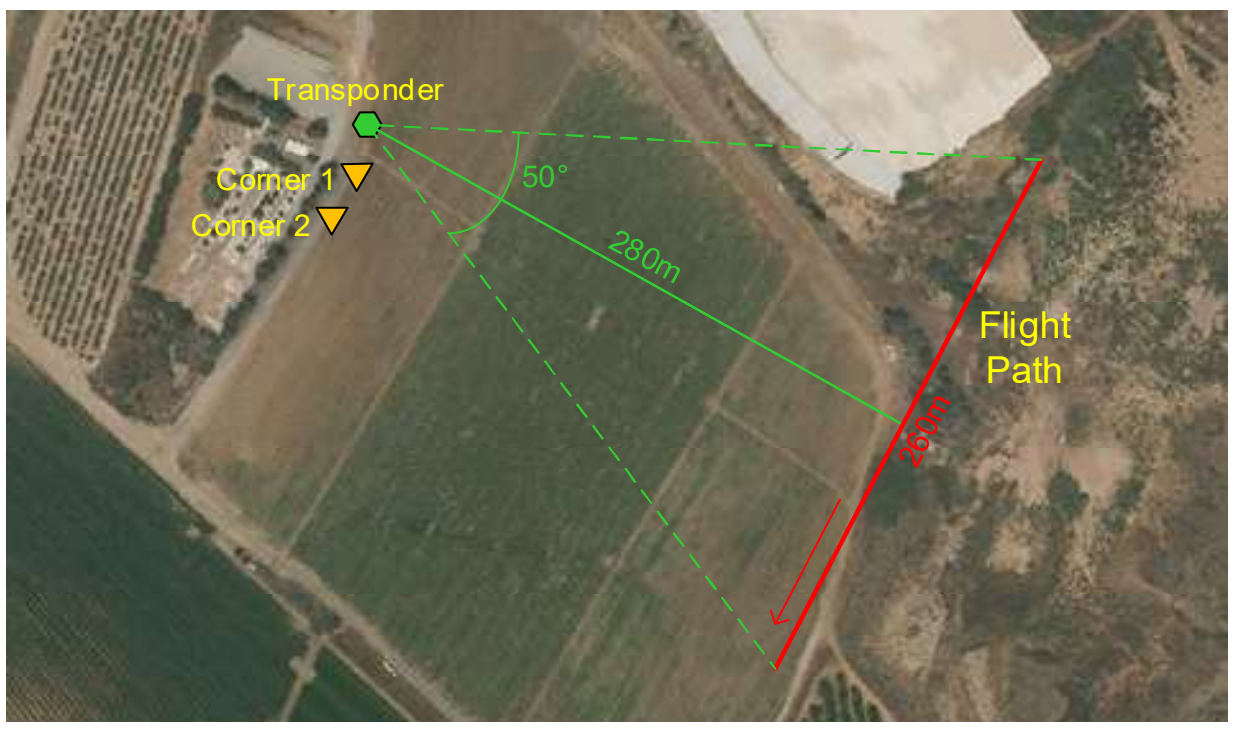

**Figure 5** Satellite view of the linear trajectory experiment setup. The drone's flight path and the ground locations of the transponder and corner reflectors are indicated.

Following data acquisition, the range to both the transponder and the corner reflector were estimated using the processing scheme described previously.

**Figure 6** shows an FFT spectrogram of the IF spectrum. The response from the corners is located at a low frequency, where it is embedded within the strong ground clutter. In contrast, the two tones from the transponder are shifted by approximately 20 MHz and 25 MHz into a clutter-free spectral region, limited only by the radar's noise floor. The transponder's response appears much wider than that of the corner reflector, as the transponder's sampling window is cropped to match its limited bandwidth, while the corner reflector's signal was analyzed over all available samples.
The range estimated for both the transponder and the corner reflector is presented in **Figure 7**. The measured range from both the absolute (Scheme 1) and relative (Scheme 2) methods are plotted in **Figure 7a**, while **Figure 7b** quantifies the measurement error using a moving standard deviation calculated over 100 consecutive pulses. For absolute range estimation (Scheme 1), the corner reflector demonstrates substantially higher precision, with a range error of 15 mm compared to 200 mm for the transponder. In contrast, when using the relative estimation scheme (Scheme 2), both the transponder and the corner reflector achieve a remarkably similar and highly precise relative range error of approximately 1 mm.

## 4.2 Circular Flight Experiment

In the second experiment, the transponder was equipped with the omnidirectional dipole antenna. The drone executed a circular flight path with a 200 m radius around the transponder at an altitude of approximately 100 m above ground. As illustrated in **Figure 8**, this trajectory allowed the transponder to be observed continuously over a full 360° azimuth range. To provide a direct performance baseline, an assembly of four 0.85 m corner reflectors was placed within the imaging scene.

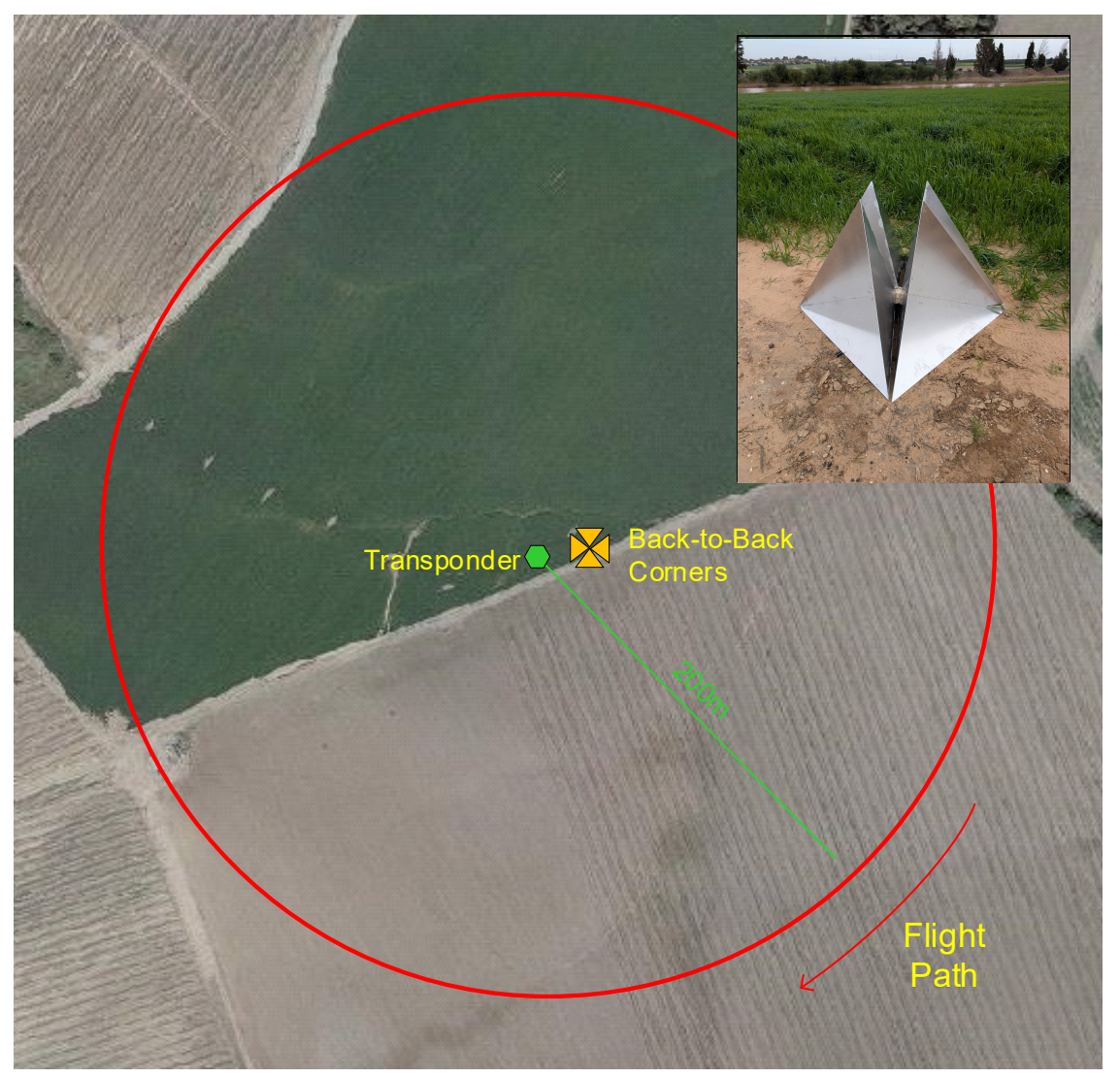

**Figure 8** Satellite view of the circular trajectory experiment setup. The inset photograph displays the four back-to-back corner reflectors deployed in the field.

The data acquired during the circular flight was processed using the same methodology detailed for the linear trajectory experiment, with the results now evaluated over a broader azimuth viewing angle range of $\pm 100°$ . **Figure 9** presents the FFT spectrogram of the resulting IF spectrum. The spectrogram clearly highlights the physical limitations of the passive reference target, with the response exhibits gaps, dropping to zero at approximately $-80°$ and $+10°$. Conversely, the transponder successfully demonstrates its omnidirectional capability, yielding a continuous response

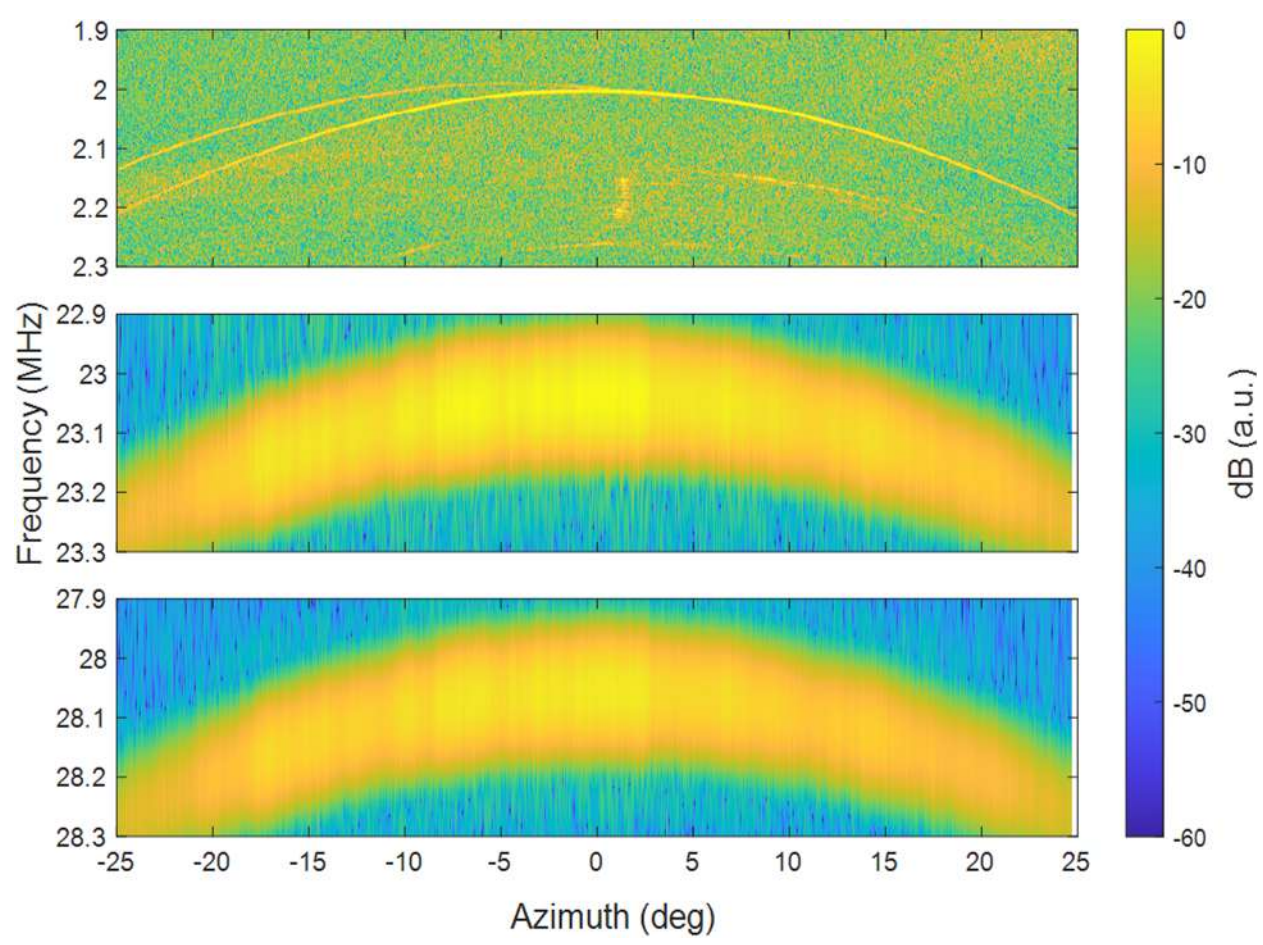

**Figure 6** Linear trajectory: Spectrogram of the IF spectrum comparing the responses from the corner reflectors (Top) and the transponder (Center and Bottom).

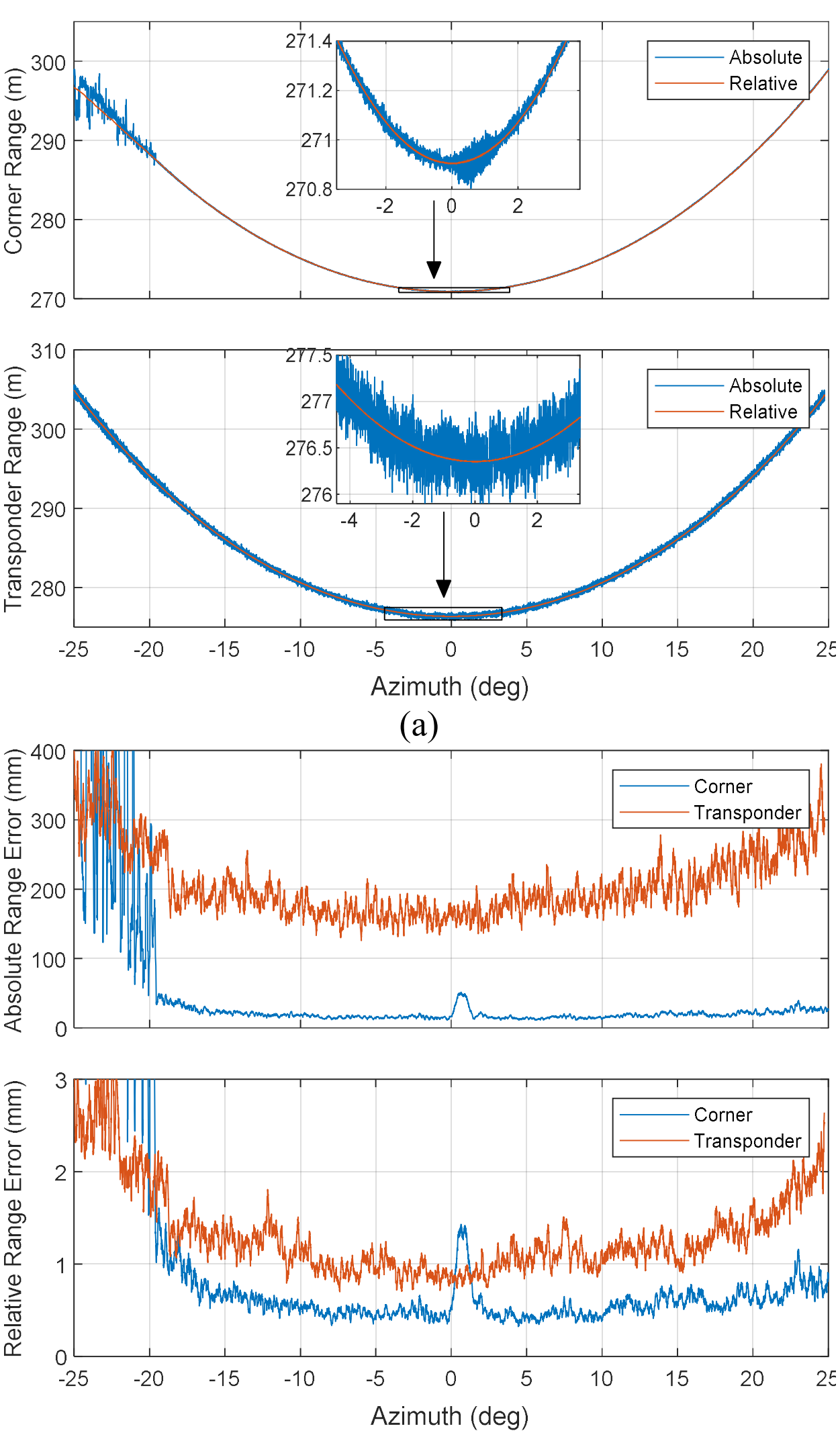

**Figure 7** Linear trajectory: Range estimation performance for the transponder and corner reflector. **(a)** Measured range over time for both absolute and relative schemes. **(b)** Corresponding measurement error.

and maintaining a constant signal-to-noise ratio across the entire azimuth sweep.

The range estimation and its associated error for both targets are detailed in **Figure 10**. While the corner reflectors achieve a substantially higher level of precision, their range estimation exhibits notable gaps. In contrast, the transponder successfully provides a continuous range deduction, even with the low-gain dipole antenna utilized.

# 5 Conclusions

In this paper, we presented and experimentally validated a novel frequency-translating transponder architecture that successfully circumvents the size constraints and narrow viewing angles associated with traditional reference targets.

The field experiment provided a direct comparison against a conventional corner reflector and highlighted a key performance trade-off originating from the transponder's intentionally narrow bandwidth. Our transponder achieved comparable relative range precision while completely eliminating the viewing angle constraints inherent to traditional passive targets.

Although the prototype was constructed with COTS components, a production version could achieve a physical volume potentially less than 1/500th that of the corner reflector used in this study, while still providing this high-precision relative range capability.

# 6 Acknowledgments

The authors would like to thank Reuven Mazor for his assistance with the field experiments, Vadim Asnin for his help in debugging the data processing, Eli Yadin for reviewing this manuscript, and Denis Kogan for his logistical support.

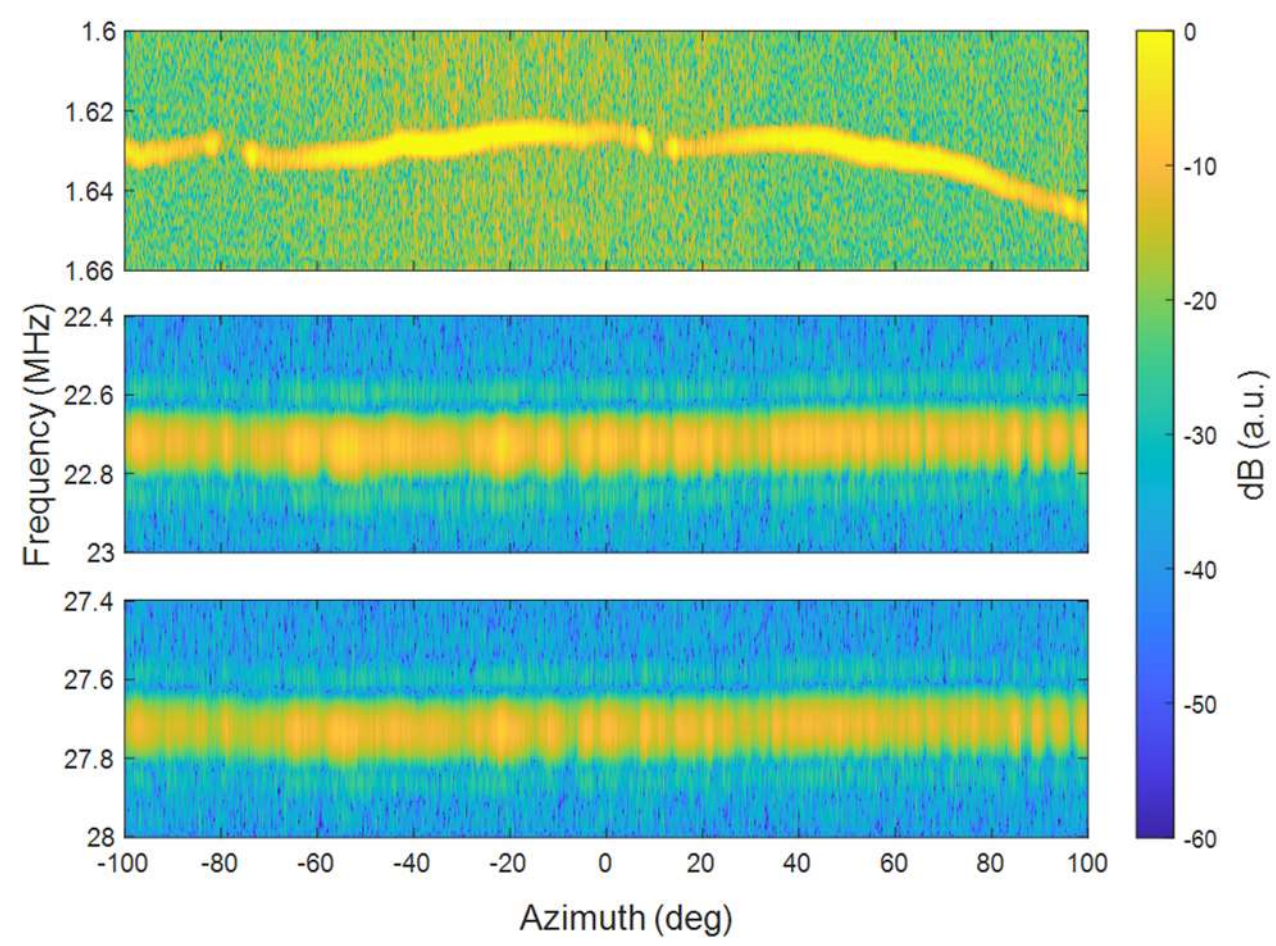

**Figure 9** Circular trajectory: Spectrogram of the IF spectrum comparing the responses from the corner reflectors (Top) and the transponder (Center and Bottom).

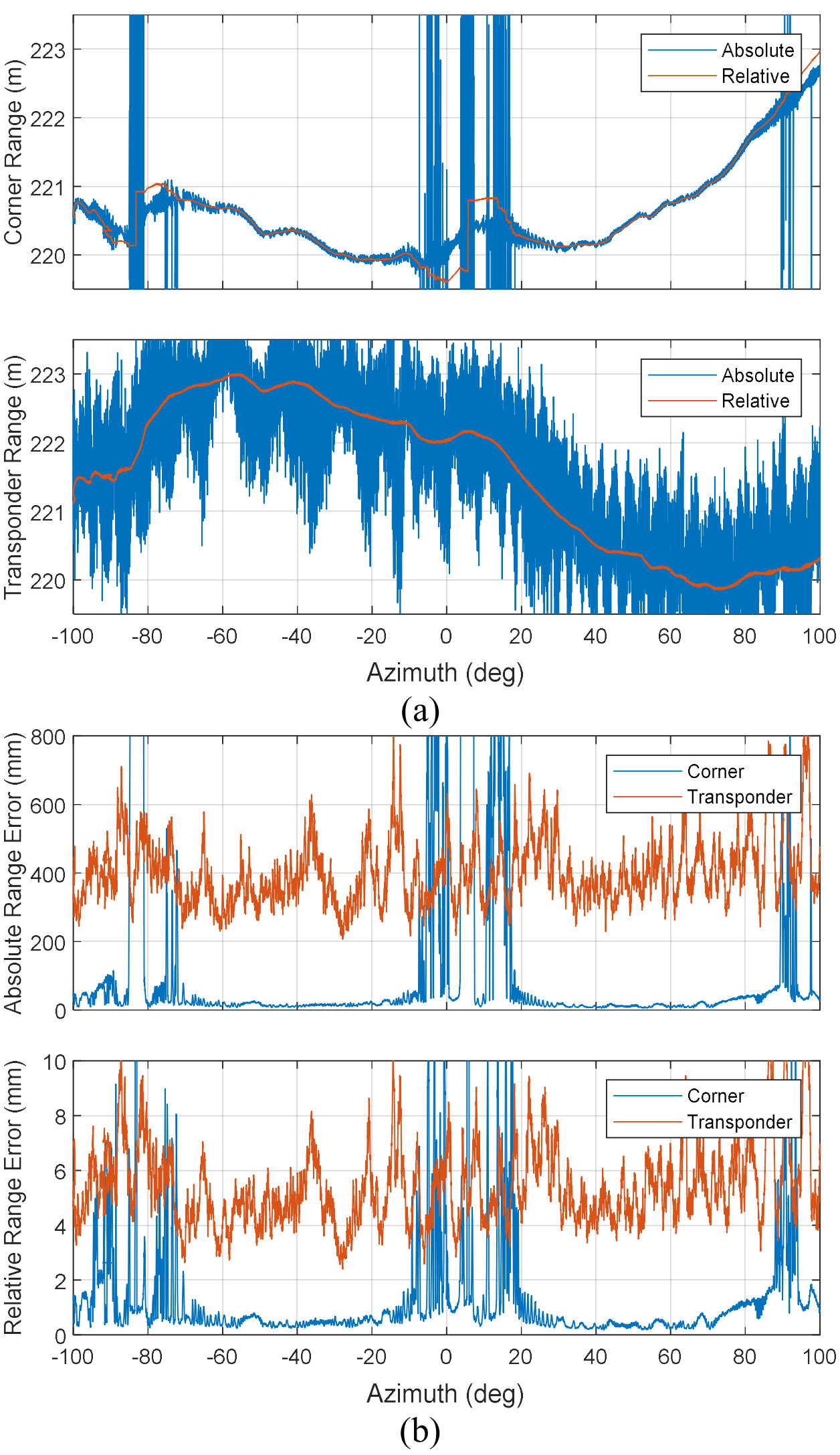

**Figure 10** Circular trajectory: Range estimation performance for the transponder and corner reflector. **(a)** Measured range over time for both absolute and relative schemes. **(b)** Corresponding measurement error.